\newcommand{\HAL}[1]{#1}
\newcommand{\NON}[1]{}

\def\ggf{\leavevmode\hbox{\raise.4ex\hbox{%
	$\scriptscriptstyle<\!<\;$}}}
\def\gdf{\leavevmode\hbox{\raise.4ex\hbox{%
	$\scriptscriptstyle\;>\!>\;$}}}
\def\leftnote#1{\leavevmode\vadjust{\setbox1=\vtop{\hsize 20mm
	\parindent=0pt\small\baselineskip=9pt
	\rightskip=4mm plus 4mm#1}
	\hbox{\kern-2cm\smash{\box1}}}}

\documentclass[twoside\HAL{,a4paper}]{article}
\usepackage[T1]{fontenc}
\usepackage[latin1]{inputenc}
\usepackage[french]{babel}
\usepackage{listings,color}
\NON{\usepackage{perpi2006}}
\HAL{\usepackage{a4wide}}
\usepackage{euler,palatino}
\usepackage{epsfig,shadow,url}

\usepackage{amsmath}
\usepackage{amssymb}

\def\figext{\HAL{pstex_t}\NON{pdftex_t}}

\begin{document}

\parindent=0pt

\title{\Large\bf Caractéristiques arithmétiques des processeurs graphiques}
\NON{\shorttitle{Caractéristiques arithmétiques des processeurs graphiques}}

\author{Marc Daumas, Guillaume Da Graça et David Defour%
\HAL{
  \\[12pt]
  DALI-LP2A (Universit\'e de Perpignan)\\
  52 avenue Paul Alduy --- 66860 Perpignan --- France\\
  \{marc.daumas, guillaume.dagrac, david.defour\}@univ-perp.fr\\[6pt]
  LIRMM  (CNRS, Université de Montpellier 2)\\
  161 rue Ada --- 34392 Montpellier Cedex 5 --- France\\
  marc.daumas@lirmm.fr
}}%

\NON{
  \address{%
    DALI-LP2A (Universit\'e de Perpignan)\\
    52 avenue Paul Alduy --- 66860 Perpignan --- France\\
    \{marc.daumas, guillaume.dagrac, david.defour\}@univ-perp.fr\\[6pt]
    LIRMM  (CNRS, Université de Montpellier 2)\\
    161 rue Ada --- 34392 Montpellier Cedex 5 --- France\\
    marc.daumas@lirmm.fr
  }
}

\HAL{\date{}}

\maketitle

\HAL{\newcommand{\Resume}[1]{\begin{abstract}#1\end{abstract}}}
\Resume{%
  Les unités graphiques ({\em Graphic Processing Units} --- GPU) sont
  désormais des processeurs puissants et flexibles. Les dernières
  générations de GPU contiennent des unités programmables de traitement
  des sommets ({\em vertex shader}) et des pixels ({\em pixel shader})
  supportant des opérations en virgule flottante sur 8, 16 ou 32 bits.
  La représentation flottante sur 32 bits correspond à la simple précision de la
  norme IEEE sur l'arithmétique en virgule flottante (IEEE-754). Les
  GPU sont bien adaptés aux applications avec un fort parallélisme de
  données.  Cependant ils ne sont que peu utilisés en dehors des
  calculs graphiques ({\em General Purpose computation on GPU} ---
  GPGPU).  Une des raisons de cet état de faits est la pauvreté des
  documentations techniques fournies par les fabricants (ATI et
  Nvidia), particulièrement en ce qui concerne l'implantation des
  différents opérateurs arithmétiques embarqués dans les différentes
  unités de traitement. Or ces informations sont essentielles pour
  estimer et contrôler les erreurs d'arrondi ou pour mettre en
  {\oe}uvre des techniques de réduction ou de compensation afin de
  travailler en précision double, quadruple ou arbitrairement étendue.
  Nous proposons dans cet article un ensemble de programmes qui
  permettent de découvrir les caractéristiques  principales des GPU en ce qui concerne l'arithmétique à virgule flottante.
  Nous donnons les résultats obtenus sur deux cartes graphiques
  récentes: la Nvidia 7800GTX et l'ATI RX1800XL.
}%

\NON{\MotsCles{GPGPU, GPU, arithmétiques virgule flottante, précision.}}

\section{Introduction et présentation des unités graphiques (GPU)}

Les unités graphiques (GPU) sont des unités spécialisées dans le
calcul intensif et régulier qui développent une puissance de calcul
bien supérieure à celle disponible sur les processeurs généralistes
\cite{Pharr2005}.  Avec l'arrivée des GPU de dernière génération, il devient 
intéressant d'utiliser ces processeurs GPU pour faire des calculs
généralistes (GPGPU) \cite{Man05}~\footnote{Voir aussi le site
  \url{http://www.gpgpu.org/}.}.  Les GPU deviennent alors des
processeurs spécialisés pour des applications régulières et à fort parallélisme de
données \cite{LegRob03}.


Après une présentation des unités graphiques (section \ref{sec:pipeline_graphique}) et des différentes
implantations de l'arithmétique à virgule flottante (section \ref{sec:virgule_flottante}), cette introduction
se termine par un état de l'art des travaux liés aux tests des propriétés flottantes d'un système (section \ref{sec:etat_art}). Nous présentons ensuite nos algorithmes, leurs implantations et les
résultats obtenus (section \ref{sec:resultat}).


\subsection{Modèle de fonctionnement des cartes graphiques: le pipeline graphique}
\label{sec:pipeline_graphique}

\begin{figure}
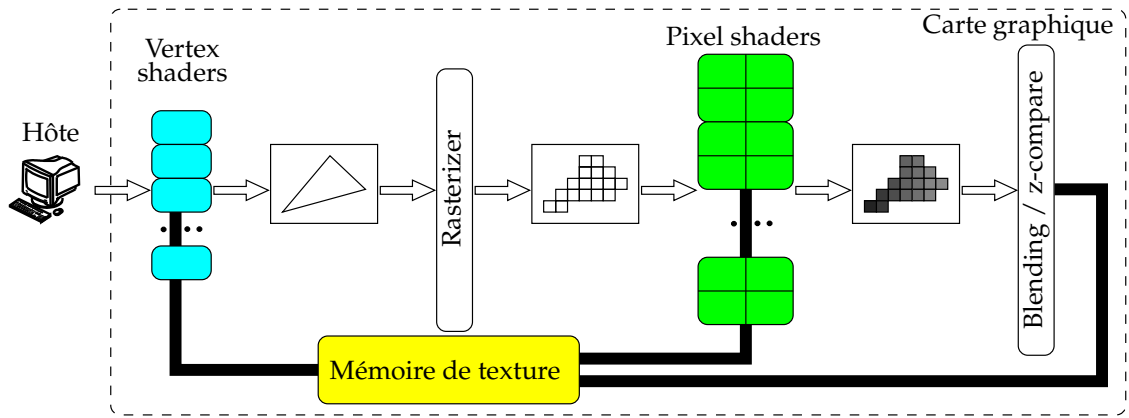

  \begin{center}
    \input \NON{../Fig/}graphics_pipeline.\figext
  \end{center}
  \caption{Vue d'ensemble du pipeline graphique}
  \label{fig:graphics_pipeline}
\end{figure}

Les GPU traitent principalement des objets géométriques et des pixels.
Les images sont créées en appliquant des transformations géométriques
aux sommets et en découpant les objets en fragments ou pixels.  Les
calculs sont réalisés par différents étages de ce que l'on appelle le
pipeline graphique, comme présenté à la
figure~\ref{fig:graphics_pipeline}. La machine hôte envoie des sommets
pour positionner dans l'espace des objets géométriques primitifs (polygones, lignes,
points). Ces objets primitifs subissent des transformations (rotations, translation, illumination\ldots) avant d'être assemblés pour créer un objet plus complexe. Ces opérations sont réalisées dans l'unité de traitement des sommets ({\em vertex shader}).

Quand un objet a atteint sa position, sa forme et son éclairage
finals, il est découpé en fragments ou pixels. Une interpolation est
effectuée pour obtenir les propriétés de chaque pixel. Les pixels sont
ensuite traités par l'unité de traitement des pixels ({\em pixel
  shader}) qui n'effectue pas des transformations géométriques mais des
tâches d'affichage comme par exemple appliquer une texture ou calculer
la couleur d'un pixel.

Le pipeline effectivement implanté diffère légèrement de celui
représenté dans la figure~\ref{fig:graphics_pipeline}. Selon les
cartes et les circuits, les constructeurs déplacent, partagent,
dupliquent ou ajoutent certaines ressources. La figure montre les
différentes étapes sur l'exemple d'un triangle.  Les vertex shaders traitent 3 sommets alors que les pixel shaders  traitent 17 pixels. Pour une figure donnée, le
nombre de pixels est presque toujours plus important que le nombre de
sommets et les architectures modernes contiennent plus d'unités de
traitement des pixels que d'unités de traitement des sommets. Le ratio
actuel est par exemple de 24 pour 8.



\begin{figure}
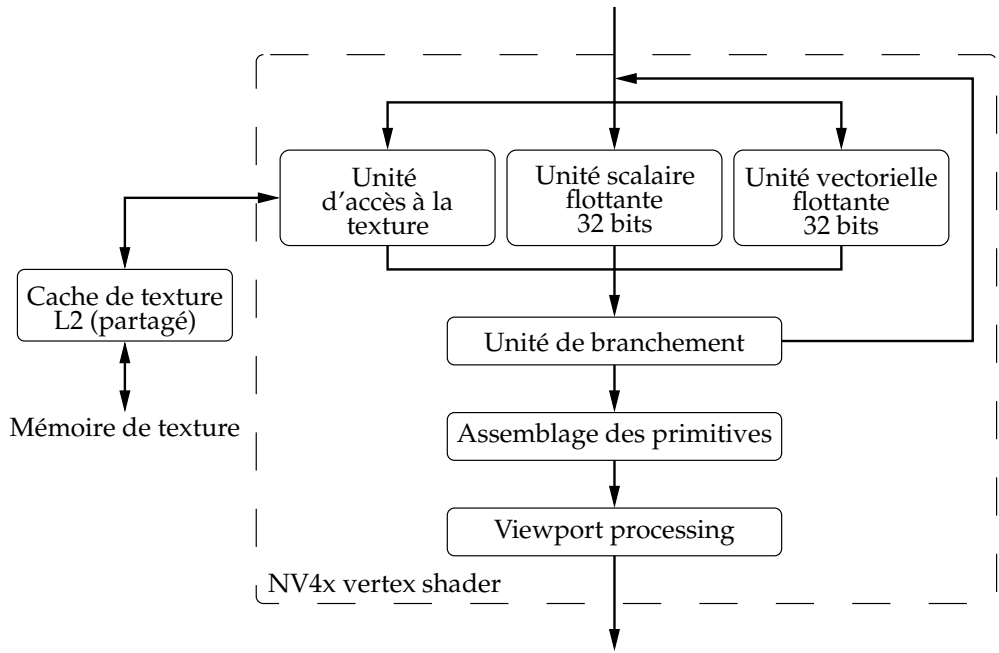

  \begin{center}
    \input \NON{../Fig/}vertexshader.\figext
  \end{center}
  \caption{Détails d'un vertex shader présent dans la Nvidia 7800GTX}
  \label{fig:vertexshader}
\end{figure}

La figure \ref{fig:vertexshader} reprise de \cite{Pharr2005} présente
l'un des 8 vertex shaders de la Nvidia 7800 GTX. Dans la
classification de Flynn \cite{Fly66,Fly72}, les 8 shaders fonctionnement en
mode MIMD ({\em Multiple Instruction Multiple Data}) entre eux.  Chaque vertex
shader est capable d'initier à chaque cycle une opération {\em Multiply
  and Accumulate} (MAD) sur 4 triplés dans l'unité vectorielle et une
opération {\em special} dans l'unité scalaire.  Les opérations  {\em special}
implantées sont les fonctions exponentielles (exp, log), trigonométriques (sin, cos) et deux fonctions inverses ($1/x$ et $1/\sqrt{x}$). Avec l'arrivée de la
version 3 du support matériel Direct~3D, les shaders sont capables d'accéder à la mémoire de texture
grâce à une unité dédiée.

\begin{figure}
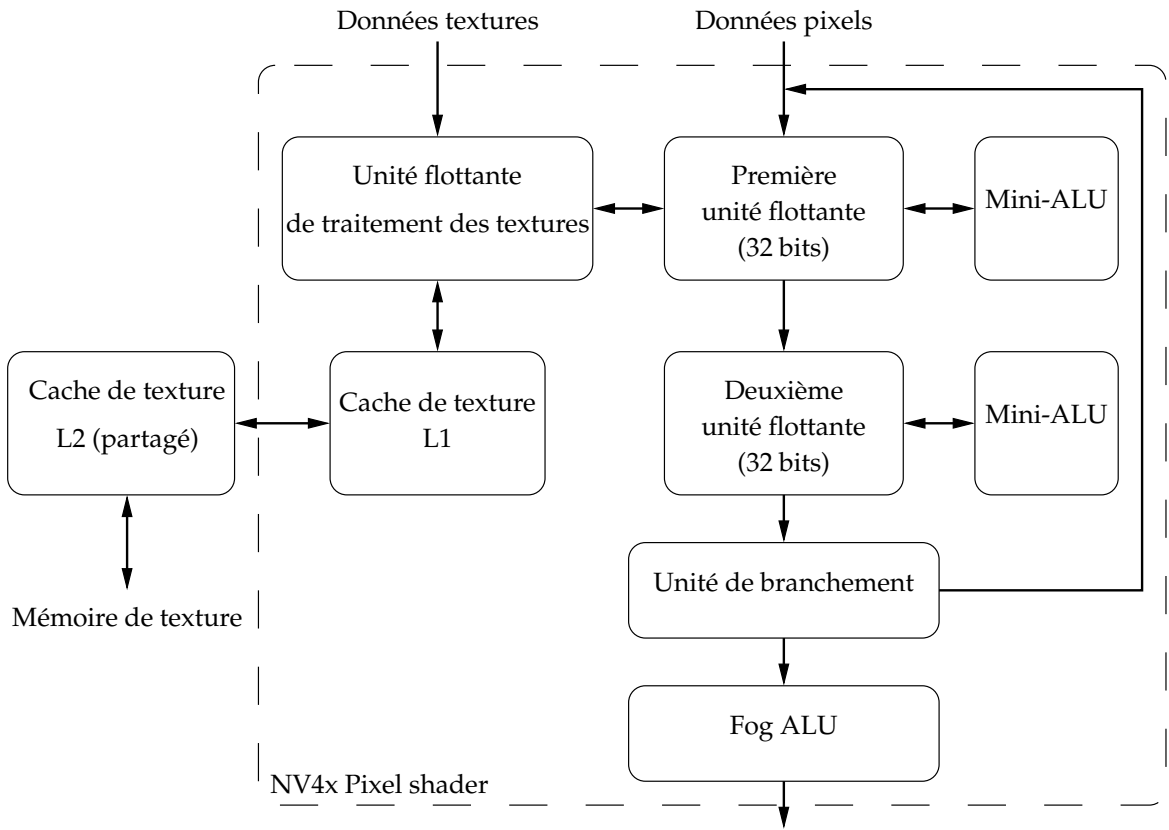

  \begin{center}
    \input \NON{../Fig/}pixelshader.\figext
  \end{center}
  \caption{Détails d'un pixel shader présent dans la Nvidia 7800GTX}
  \label{fig:pixelshader}
\end{figure}

La figure~\ref{fig:pixelshader} décrit l'un des 24 pixel shaders de la
Nvidia 7800 GTX.  Les 24 shaders fonctionnent en mode SIMD ({\em Single
  Instruction Multiple Data}) entre eux. La première unité flottante exécute
selon le programme 4 MAD ou un accès à la texture via
l'unité de traitement des textures.  Le résultat est envoyé à la
deuxième unité flottante qui exécute 4 MAD.  Dans le cas
de la Nvidia 7800 GTX, chaque pixel shader dispose d'une mémoire dédiée appelée cache
de texture de niveau 1  et d'une unité capable de premiers traitements
sur les textures.

\subsection{Implantation sur GPU de l'arithmétique à virgule flottante en regard des normes ANSI et ISO}
\label{sec:virgule_flottante}
Les GPU travaillent sur différents formats de représentation des
nombres. L'arithmétique à virgule flottante offre un spectre de
nombres représentables plus étendu que l'arithmétique entière ou à
virgule fixe sans que le programmeur ait besoin de gérer manuellement
décalages et recadrages. L'implantation de l'arithmétique à virgule
flottante \cite{Dau05} est normalisée par deux normes américaines en cours de
révision\footnote{Voir le site \url{http://grouper.ieee.org/groups/754/}.}
IEEE-ANSI 754 \cite{Ste.81,Ste.87} et 854 \cite{CodKar.84} et par une
norme internationale IEC-ISO 60559 \cite{IEC89}. Des implantations
très différentes de l'arithmétique à virgule flottante existent
\cite{BolDau04a} mais elles ne sont pas utilisées par les GPU.

\begin{table}
  \caption{Format de représentation des nombres à virgule flottante sur GPU et CPU}
  \label{fpformat}
  \begin{center}
    \begin{tabular}{|l|c|c|c|c|c|}\hline
      Référence  & \multicolumn{4}{|c|}{Nombre de bits}                              & Valeurs non   \\ \cline{2-5}
                 & \parbox{2cm}{\centering Total} & \parbox{2cm}{\centering Signe} & \parbox{2cm}{\centering Exposant} & \parbox{2cm}{\centering Fraction} & numériques    \\ \hline \hline
      Nvidia     & 16 & 1 & 5  & 10 & NaN, Inf        \\ \cline{2-5}
                 & 32 & 1 & 8  & 23 &                 \\ \hline
      ATI        & 16 & 1 & 5  & 10 & Absentes        \\ \cline{2-5}
                 & 24 & 1 & 7  & 16 &                 \\ \cline{2-6}
                 & 32 & 1 & 8  & 23 & Non documentées \\ \hline \hline
      ANSI-ISO   & 32 & 1 & 8  & 23 & NaN, Inf        \\ \cline{2-5}
                 & 64 & 1 & 11 & 52 &                 \\ \hline
    \end{tabular}
  \end{center}
\end{table}

Les normes précédentes représentent un nombre par trois champs, une
mantisse $m$, un exposant $e$ et un signe $s$, la base de l'exposant
est fixée à 2 ou à 10. Les GPU utilisent tous la base 2. La mantisse
est un nombre en virgule fixe avec 1 bit avant la virgule, l'exposant
est un entier biaisé. Les normes précédentes imposent que le bit de
la mantisse à gauche de la virgule soit égal à 1 ce qui est toujours
possible sauf si le nombre à représenter est trop petit. On parle selon les cas de représentation normalisée ou dénormalisée.
Pour économiser de la mémoire, on ne stocke pas ce premier bit de la mantisse égal à 1 pour les nombres normalisés et on appelle
fraction $f$ les bits de la mantisse à droite de la virgule.  Ainsi,
un nombre flottant normalisé $x$, s'écrit
$$
x ~ = ~ (-1)^s ~ \cdot ~ m ~ \cdot ~ 2^e ~ = ~ (-1)^s ~ \cdot ~ 1,f ~ \cdot ~ 2^e.
$$

Les normes définissent deux formats présentés dans le
tableau~\ref{fpformat}. Le format simple précision qui est codé sur 32
bits et le format double précision qui est codé sur 64 bits. Pour
traiter les dépassements de capacité ou les situations
exceptionnelles (division par zéro), les normes définissent les valeurs non numériques que
sont les infinis ou les non-nombres ({\em Not a Number} --- NaN).

Le résultat d'une opération arithmétique ($+$, $\times$, $/$,
$\sqrt{~}$) n'est presque jamais représentable bien que les opérandes
sont représentables. Il faut l'arrondir.  Les normes imposent que
l'on utilise un des quatre modes d'arrondi prédéfinis: par défaut, par
excès, tronqué, ou au plus près. Des travaux anciens \cite{Coo78}
montrent que l'on peut arrondir précisément les quatre opérations
usuelles précédentes, en utilisant uniquement trois bits
supplémentaires par rapport au format du résultat. C'est-à-dire comme si les opérations intermédiaires
avaient été infiniment précises. On parle des  bits {\em guard}, {\em
  round} et {\em sticky}.

Une règle d'arrondi au plus proche ne définit pas un résultat unique
quand le résultat intermédiaire exact se situe à égales distances
entre deux nombres représentables.  Pour des raisons statistiques, l'arrondi normalisé favorise le
nombre dont la mantisse se termine par un 0 et on parle d'arrondi pair
\cite{ReiKnu75}.

Le tableau~\ref{fpformat} reprend les formats implantés sur les cartes
graphiques que nous avons étudiées. Les données concernant le support pour les valeurs non numériques sont issues de \cite{Cebenoyan2005}. Le format sur 32 bits est lié à la
version 3.0 de Direct~3D.  Ce format s'inspire du format simple
précision des norme ANSI-IEEE et ISO-IEC, mais les constructeurs ne garantissent
pas la compatibilité de leur unité avec ces normes. La documentation disponible sur les cartes graphiques ne permet pas de déterminer en quoi ces implantations diffèrent de la norme. Il est donc nécessaire de mieux
comprendre le fonctionnement de l'arithmétique flottante sur GPU pour
maîtriser le comportement numérique des applications scientifiques exécutées sur GPU.

\subsection{État de l'art et travaux antérieurs sur CPU et sur GPU}
\label{sec:etat_art}

Les premiers logiciels réalisés pour tester l'arithmétique à virgule
flottante sur CPU avaient pour unique but de découvrir les
caractéristiques des fonctionnalités implantées
\cite{Cod88,GenMar74,Sch81}. Suite à l'adoption  massive de la norme
IEEE-754, certains logiciels sont apparus pour vérifier la bonne implantation de cette
norme par une série de tests bien choisis
\cite{Kar85,Par99,VerCuyVer01a} alors que d'autres proposaient
d'implanter et de tester des fonctions élémentaires, spéciales ou
complexes \cite{Cod93a,Cod93b}.
Certains logiciels tels UCBTest\footnote{Ce programme est disponible sur le dépôt Netlib
  \url{http://www.netlib.org}.} testent à la fois la conformité des
fonctionnalités normalisées, la qualité des autres fonctionnalités
usuelles et le bon fonctionnement de la chaîne de  compilation.




Comme nous l'avons évoqué  à la section précédente, les GPU ne respectent pas la norme IEEE-754 contrairement à la
majorité des CPU.  Certaines  hypothèses sur le comportement arithmétique des opérateurs flottants doivent être vérifiées avant
de pouvoir porter un programme pour CPU sur un GPU comme, par exemple, émuler des opérateurs arithmétiques en précision multiple \cite{Graca2006}.


{\em Paranoia} \cite{Kar85} est un outil qui teste certaines
propriétés de l'arithmétique flottante des CPU. Un sous-ensemble de
ces tests a été adapté pour pouvoir être exécuté sur certains GPU
\cite{Hillesland2004}. \NON{Nous publions en annexe de notre rapport
  de recherche\footnote{Bientôt disponible sur HAL-CCSD
    \url{http://}.} les résultats complets du test de la carte
  graphique Nvidia 7800GTX.} \HAL{Les résultats complets du test de la
  carte graphique Nvidia 7800GTX sont publiés en annexe.} En plus des
binaires du logiciel, les auteurs fournissent un tableau de synthèse
sur l'ATI R300 \cite{Hillesland2004}. À ce jour, ce programme ne
fonctionne pas sur notre ATI RX1800XL.
%
Ses sources ne sont pas disponibles ce qui nous
empêche d'aller plus loin que les seuls messages générés par
son exécution.  Par exemple, nous ne pouvons pas déterminer
quels opérateurs ont été testés parmi tous les opérateurs disponibles
dans les vertex et pixel shaders et quels vecteurs de test ont été
utilisés au delà de {\em Paranoia}.

Certaines caractéristiques se dégagent de ces premiers tests:
\begin{itemize}
\item L'addition et la multiplication sont tronquées sur les deux GPU.
\item La soustraction semble bénéficier d'un bit de garde sur Nvidia
  mais pas sur ATI.
\item La multiplication implante un arrondi fidèle sur les deux GPU \cite{Pri92}.
\item L'erreur de division semble indiquer que la division est
  implantée par la multiplication du dividende par une approximation
  de l'inverse du diviseur \cite{BriMulRai04}.
\end{itemize}

En conclusion, les travaux initiés montrent que de nombreuses
questions restent sans réponse et que les réponses ne pourront être
apportées que par un logiciel dont les sources sont disponibles. C'est
le cas pour les logiciels cités précédemment et ciblant les CPU et pour les implantations des algorithmes décrits dans
la suite qui sont disponibles sur demande auprès des auteurs pour évaluation.


\section{Découverte des caractéristiques des unités arithmétiques des GPU}
\label{sec:resultat}
Nous avons défini et implanté des algorithmes pour mieux comprendre le
fonctionnement de l'addition, de la multiplication et le stockage des
nombres à virgule flottante dans les registres et en mémoire. Bien que
ces algorithmes peuvent être adaptés à d'autres formats de données, nous
avons ciblé le format simple précision sur 32 bits. Les algorithmes
ont été écrits dans une version préliminaire en OpenGL et utilisent les
{\em Frame Buffer Object} pour le stockage dans les textures. Ces
premiers programmes fonctionnent parfaitement sur la carte Nvidia 7800
GTX avec le driver ForceWare 81.98.  En revanche ils ne fonctionnent
pas sur la carte ATI RX1800XL et le driver Catalyst 6.3 et nous avons
réécrit les programmes en DirectX.  Nos algorithmes sont ainsi
disponibles pour OpenGL et DirectX.



\subsection{Stockage des nombres à virgule flottante dans les registres et en mémoire}


Le GPU est souvent utilisé comme coprocesseur. Des données sont
générées dans le CPU pour être transférées dans la mémoire de texture avant d'être traitées par le GPU.
Pour savoir si des conversions interviennent lors des transferts du
CPU vers le GPU, nous avons envoyé des nombres dénormalisés, des NaN
et des nombres infinis du CPU vers le GPU pour ensuite les récupérer
sur le CPU. Nous avons aussi effectué des opérations sur GPU générant
ces valeurs spéciales et récupéré le résultat sur CPU.

Les résultats obtenus montrent que lors d'un transfert sans aucune
opération, les nombres dénormalisés sont remplacés par 0 et les
infinis ne sont pas modifiés sur Nvidia ou ATI. Les NaN ne sont pas
modifiés sur Nvidia mais ATI transforme les sNaN ({\em signaling NaN})
en qNaN ({\em quiet NaN}).

Sur certaines architectures, les registres internes stockent les nombres avec une
précision supérieure à celle des données en mémoire \cite{Int97a} et avec une
dynamique plus grande pour l'exposant \cite{Mar2K}. La conversion vers le format final est réalisée lors de l'écriture en mémoire. Nous avons testé si certaines unités
des GPU se comportent de manière similaire. Pour tester l'exposant
maximum, nous avons utilisé des vecteurs pour calculer
$$(\text{MAX\_FLOAT} + \text{MAX\_FLOAT}) - \text{MAX\_FLOAT}$$
où $\text{MAX\_FLOAT}$ est le plus grand nombre représentable au format
simple précision. Nous avons vérifié le nombre de bits de la mantisse en utilisant des vecteurs pour calculer
$$(1,5 - 2^{-i}) + 2^{-i}$$
pour $i$ variant de 1 à 64.

Nous avons voulu savoir si les différents MAD, où $\text{MAD}(x, y,
z)$ effectue l'opération $x \times y + z$, conservent dans l'accumulateur plus de bits de
mantisse du produit $x \times y$ que la précision de travail pour les
utiliser quand les chiffres de poids forts du produit sont compensés
par le troisième opérande $z$. C'est le cas des architectures
normalisées récentes qui implantent un FMA en arithmétique à virgule
flottante \cite{Mar2K,Fre05}.  Nous avons généré des vecteurs de tests
aléatoires et comparé le reste de la multiplication sur GPU avec la
valeur exacte calculée sur CPU,
$$\text{MAD} (x, y, - x \otimes y)$$
où $x \otimes y$ est la multiplication arrondie au plus près au format
simple précision sur CPU.





L'exécution des tests montre qu'aucun des registres temporaires des
vertex et pixel shaders des cartes ATI et Nvidia n'utilise une plage
d'exposant étendue pour éviter les débordements ou un nombre de bits de
mantisse supérieur pour augmenter la précision des calculs. Par
ailleurs, aucun des MAD ne conserve le produit sur une précision
étendue au delà des 24 bits de précision des nombres flottants simple
précision.

\subsection{La multiplication}

\begin{figure}
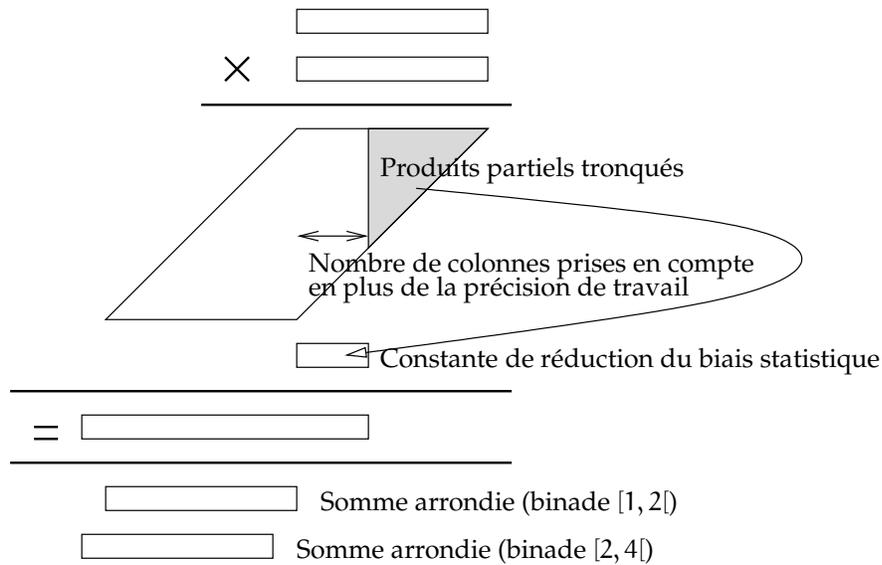

  \begin{center}
    \input \NON{../Fig/}Mul_1.\figext
  \end{center}
  \caption{Produits partiels tronqués pour accélérer le multiplieur et réduire sa taille}
  \label{fig:produits_partiels}
\end{figure}

Les résultats de l'exécution de {\em Paranoia} nous laissent penser que
la multiplication est tronquée à la fois sur ATI et sur Nvidia. Nous
avons testé la façon dont cette troncature était effectuée à
l'intérieur des pixel shaders. Pour cela nous avons souhaité
déterminer si tous les produits partiels était générés ou si pour une
question de rapidité, les bits de poids faibles de certains produits
partiels étaient ignorés comme présenté à la
figure~\ref{fig:produits_partiels} \cite{SchSwa93b}. Dans ce cas, une constante est
ajoutée à la somme des produits partiels pour réduire  le biais
statistique introduit par la troncature d'une quantité toujours
positive.

Nous avons multiplié deux vecteurs de 24 nombres: un premier vecteur composé uniquement de
la valeur $2^{23}+3$ et un deuxième vecteur défini par la récurrence suivante pour $i > 0$
$$
    y_1      = 1                  ~~~~~~~~~~ 
    y_{2i}   = y_{2i - 1} + 1     ~~~~~~~~~~
    y_{2i+1} = 4 y_{2i - 1} + 2.
$$
Cette suite peut aussi s'interpréter à l'aide des opérateurs binaires \emph{ou logique} ($|$) et 
\emph{décalage vers la gauche} ($<<$) en :
$$
    y_1      = 1                            ~~~~~~~~~~ 
    y_{2i}   = y_{2i - 1}        ~ | ~ 01_2  ~~~~~~~~~~
    y_{2i+1} = (y_{2i - 1} << 2) ~ | ~ 10_2.
$$
Ce deuxième vecteur alterne une valeur qui crée une chaîne de 1 sans
propagation dans les bits de poids faibles pour vérifier la
troncature, avec une valeur générant une propagation de retenue à
partir de la colonne la plus à droite jusqu'au premier bit du résultat
attendu.

Les résultats des tests nous indiquent que le multiplieur des pixel
shaders traite les produits partiels  jusqu'à la 9ème colonne
pour ceux d'ATI et jusqu'à la 6ème colonne pour Nvidia. Ces résultats
sont confirmés par des tests aléatoires qui montrent qu'un
biais est ajouté pour compenser les produits partiels manquants.

Un deuxième test composé de 2 vecteurs de $2^{23}$ nombres
nous donne plus de précision. Il calcule les produits
$$(2^{23} + 1) \times (2i + 1) ~~~~~ \text{pour} ~~~~~0 < i < 2^{23}.$$
Ce test détermine la valeur de la constante de
réduction du biais statistique. 





Nous avons testé sur deux  milliards de valeurs aléatoires sur ATI et
Nvidia que $A \times B$, $(-A) \times (-B)$ et $-(A \times (-B))$
sont égaux, ainsi que $(-A) \times B$ et $A \times (-B)$. Selon toute
probabilité, la notation signe-valeur absolue est utilisée et le
multiplieur calcule de façon séparée le signe du résultat et sa valeur
absolue.

\subsection{L'addition}


Lors de l'utilisation de l'outil {\em Paranoia}, nous avons remarqué
que la soustraction sur les processeurs Nvidia semblait disposer d'un
bit de garde. En l'absence d'informations supplémentaires, nous avons
cherché à clarifier le comportement de la soustraction des pixel et
vertex shaders de l'ATI et de la Nvidia. Les pixel shaders des deux
architectures testées comportent deux MAD cascadés. Nous avons déterminé
le comportement du premier additionneur en calculant
$$
1,5 - 2^{-i}
$$
pour $i$ variant de 1 à 64. Nous avons remarqué que le résultat obtenu
était égal à $1,5$ dès que $i \geq 26$ sur ATI et Nvidia dans les
vertex et pixel shaders. Ce résultat laisse penser que la première
soustraction bénéficie de deux bits de garde.

De façon similaire nous avons testé le deuxième additionneur des pixel
shaders en calculant les deux soustractions suivantes
$$
(1,5 - 2^{-i})-1.5
$$
pour $i$ variant de 1 à 64. Le résultat retourné était égal à $0$
lorsque $i \geq 25$ sur ATI et $i \geq 26$ sur Nvidia. Les deux
additionneurs cascadés des pixel shaders de la Nvidia se comportent donc
de façon similaire. En revanche, nous pouvons penser que dans le cas
où une seule addition est lancée dans les pixel shaders d'ATI, alors
elle est effectuée par le deuxième additionneur qui dispose de deux bits de
garde alors que lorsque deux additions sont lancées, les deux additionneurs
sont utilisés et le premier ne dispose que d'un seul bit de garde.

L'utilisation d'un bit de garde permet de ne pas commettre d'erreur
d'arrondi quand on calcule la différence de deux nombres proches mais
dont les exposants diffèrent d'une unité.  En revanche, la présence
d'un ou de plusieurs bits de garde supplémentaires dans une arithmétique tronquée peut
faire que la propriété suivante n'est plus vérifiée \cite{Pic76,PicVig93}
$$e(x) - e(y) > t ~~~ \Longrightarrow ~~~ x \oplus y = x$$
où $x \oplus y$ est le résultat de l'addition sur GPU, $e(\cdot)$ est
la valeur de l'exposant de la variable considérée et $t$ est le nombre
de bits de mantisse du format de travail. Cette propriété est nécessaire pour que la différence
$$x + y - x \oplus y$$
est toujours représentable en machine sauf dépassement de capacité. Il faut par la suite modifier   certains algorithmes de calcul à précision multiple 
\cite{Pri92,Graca2006}.
Nous continuons nos travaux pour comprendre
pourquoi les additionneurs des GPU testés disposent de deux bits de
garde et non pas d'un seul.


Nous avons lancé les mêmes
additions que précédemment mais en utilisant des flottants sur 16 bits
au lieu de 32. Les résultats obtenus sur Nvidia montrent que les
additions sont effectuées à précision maximale (26 bits de précision)
pour ensuite être arrondies dans le format de destination souhaité
(11 bits).

\section{Conclusion et perspectives}

Nous avons vu des algorithmes adaptés aux caractéristiques des
opérateurs flottants des processeurs graphiques.  Grâce à ces
algorithmes, nous avons pu tester les propriétés des additionneurs et
multiplieurs des vertex et pixel shaders de la Nvidia 7800 GTX ainsi
que de l'ATI RX1800XL. Nous avons, entre autre, montré que les
registres temporaires stockent les nombres flottants uniquement sur
32 bits. Nous avons également déterminé le comportement des
multiplieurs qui utilisent un biais pour compenser la troncature. Et
enfin nous avons montré que les additionneurs disposent de 2 bits de
garde ce qui explique certaines surprises lors du calcul
de l'erreur d'arrondi. Ces résultats constituent une
première étape pour la construction d'algorithmes numériques plus
précis et plus rapides sur GPU. Cependant, de nombreux tests
complémentaires restent à écrire pour caractériser précisément le
comportement de ces opérateurs et de ceux que nous n'avons pas encore
testés (fonctions trigonométriques, logarithmiques et inverses).
 

\bibliographystyle{plain}
\bibliography{alternate,daumas_ref}

\HAL{
  \appendix
  
  \section{Messages générés par Paranoia pour la Nvidia 7800GTX}
  
  {\small
    \begin{verbatim}
Created a 128x128 RenderTexture with BPP(32, 32, 32, 32)                                            
Created a 128x128 RenderTexture with BPP(32, 32, 32, 32)                                            
Created a 128x128 RenderTexture with BPP(32, 32, 32, 32)                                            
Created a 128x128 RenderTexture with BPP(32, 32, 32, 32)                                            
Created a 128x128 RenderTexture with BPP(32, 32, 32, 32)                                           
Created a 128x128 RenderTexture with BPP(32, 32, 32, 32)                                            
Created a 256x256 RenderTexture with BPP(32, 32, 32, 32) 
                                                                                                                                               
Verifying U1 U2 F9 and B9                                                                          
VERIFY U1, U2, F9, B9, FAILURE CAN ALSO MEAN INCOMPLETE CARRY PROPOGATION                           
U1 U2 F9 B9 Verified            

----BEGINNING PARANOIA TESTS----                                                                                                                                                                                                                                                                            

FUZZY TEST                                                                                          
There are two pairs of conditions checked.  One of the two must be true.                            
The first of the pair is X != 1.  The second is (X - 1/2) - 1/2 ==0                                 
No fuzziness detected in comparison                                                                                                                                                                                                                                                                         

MULTIPLICATION GUARD BIT TESTS                                                                      
This first checks that 1*x and x*1 behave the same                                                  
Then it checks if (1+U2)*2 and 2*(1+U2) behave the same                                             
Multiplication guard bit tests: passed passed passed passed                                         
MULTIPLICATION: Seems to have guard bit                                                                                                                                                                                                                                                                     

MULT ACCURACY TEST                                                                                  
Checks multiplication accuracy per line 1980                                                        
MULTIPLICATION: Accuracy tests passed                                                                                                                                                                                                                                                                       

DIVISION GUARD BIT TESTS                                                                            
The following three division guard bit tests are from line 2000                                     
DIVISION: 1/(1-U2) - (1 + U2) == 0 test: failed                                                     
DIVISION: 1/3 == 3/9 test: passed                                                                   
DIVISION: 3/9 == 9/27 test: passed                                                                  
These are tests of X/1 == 1 from line 2040 and 1/(1+U2) < 1 from 2070                               
DIVISION: F9/1 == F9 test: passed                                                                   
DIVISION: (1+U2)/1 == (1+U2) test: passed                                                           
DIVISION: 1/(1+U2) < 1 test passed                                                                  
DIVISION: FAILURE: Division lacks guard digit                                                                                                                                                                                                                                                               

ADDITION/SUBTRACTION GUARD BIT TESTS                                                                
Subtraction guard bit tests: passed passed passed passed                                            
SUBTRACTION: Seems to have guard bit                                                                                                                                                                                                                                                                        

SUBTRACTION COMPARISON TESTS                                                                        
Tests for consistency in comparison and subtraction.                                                
Specifically either (1-U1) == 0 or (1-U1) - 1 < 0.                                                  
Subtraction comparison tests (at least one of the following must pass):
	    failed passed
SUBTRACTION: Comparison and Subtraction consistent (2170)                                                                                                                                                                                                                                                   

MULTIPLICATION ROUNDING TESTS                                                                       
Tests on line 2390                                                                                    
(1.5-U2)(1+U2) = 1.5 + 0.5*U2 -U2^2 -> 1.5 test: failed                                             
(1.5+2*U2)(1-U2) = 1.5 + 0.5*U2 -U2^2 -> 1.5 test:  passed                                          
(1.5-2*U2)(1+U2) = 1.5 - 0.5*U2 - 2*U2^2 -> 1.5 - U2 test:  failed                                  
(1.5+U2)(1-U2) = 1.5 -0.5*U2 -U2^2 <= 1.5 - U2 test:  passed                                      
Multiplication rounds up when it should round down (first four tests)                               
Tests based on calculations on line 2400                                                              
(1.5+U2)(1+U2) = 1.5 + 2.5*U2 + U2^2 -> up to 1.5+3*U2 so EXACT                                     
(1.5-2*U2)(1-U2) = 1.5 - 3.5*U2 + U2^2 -> Rounded to 1.5-4*U2, so CHOPPED                           
(1.5 + 2*U2)*(1+U2) = 1.5 + 3.5*U2 + U2^2 -> up to 1.5+4*U2 so EXACT                                
(1.5-U2)*(1-U2) = 1.5 - 2.5*U2 + U2^2 -> Rounded to 1.5-3*U2, so CHOPPED                            
(1+2*U2)(1-U2) = 1 + U2 - 2*U2^2 -> Rounded to 1.0+1*U2, so NEITHER                                 
(1+U2)(1-U2) = 1-U2^2 -> up to 1 so EXACT                                                         
MULTIPLICATION: Is neither chopped nor correctly rounded                                                                                                                                                                                                                                                    

DIVISION ROUNDING TESTS                                                                             
Tests on line 2480                                                                                  
DIVISION rounding: (1.5+u2+u2)/(1+u2) - 1.5 <= 0 test: failed                                         
(1.5+u2+u2)/(1+u2) = 1.500000119 sign = 0 biased exponent = 127
		   mantisa = 400001                  
DIVISION rounding: ((1.5-U2-U2)-1.5)/(1-U2) - (1.5-U2-U2) <= 0 test:  failed                        
DIVISION rounding: ((1.5+U2+U2)+U2)/(1+U2) <= 1.5+U2 test:  failed                                  
DIVISION rounding: (1.5-U2-U2)/(1-U2) <= 1.5-U2 test:  passed                                       
DIVISION failed first 4 tests, rounding up where it should round down,                                       
and is therefore neither clamped nor correctly rounded                                     
Tests based on calculations on line 2490                                                            
DIVISION rounding (X on 2490): 1.5/(1+U2) - (1.5-U2) test: X==0 so EXACT                            
DIVISION rounding (Y on 2490): (1.5-U2)/(1-U2) test: Y==0 so EXACT                                  
DIVISION rounding (Z on 2490): (1.5+U2)/(1+U2) - 1.5 test: Z==0 so EXACT                            
DIVISION rounding (T on 2490): 1.5/(1-U2) - (1.5+U2+U2) test: T==0 so EXACT                         
DIVISION rounding: (1+U2+U2)/(1+U2) - (1+U2) == 0 test: Y2 < 0 so CHOPPED                           
DIVISION rounding: (F9-U1)/F9 -0.5 == F9-0.5 test:                                                  
Neither exact (Y1==F9-0.5) nor chopped (Y1 < F9-0.5)                                                
DIVISION: Is neither chopped nor correctly rounded                                                                                                                                                                                                                                                          

SUBTRACTION ROUNDING TESTS                                                                          
Tests on line 2620: If both are true, then Paranoia presumes chopping                                 
1-U1*U1 -> to 1, (false)                                                                            
1+U2*(0.5-U2) = 1+0.5*U2 - U2^2 -> 1 (true)                                                       
Tests on line 2650.  Failure means not correctly rounded.                                             
1+(0.5+U2)*U2 = 1+0.5*U2 + U2^2 -> 1 (not correct)                                                  
1+(0.5-U2)*U2 = 1 +0.5*U2 -U2^2 -> 1 (correct)                                                    
Tests on line 2670.  Failure means not correctly rounded.                                             
1-(0.5+U2)*U1 = 1 - 0.5*U1 - U2*U1 -> F9 (correct)                                                  
1 - (0.5-U2)*U1 = 1 - 0.5*U1 + U2*U1 -> 1 (correct)                                               
S = (X+Y) + (Y-X) == 0 test (line 2720): passed                                                     
SUBTRACTION: Is neither chopped nor correctly rounded                                                                                                                                                                                                                                                       

SUBTRACTION ROUNDING TESTS - LESS MACs                                                              
This is the same set of subtraction rounding tests, but                                             
some effort has been put into getting rid of some                                                   
multiply - accumulates.  This is done by computing operands                                         
on the CPU, so if there is a serious problem with GPU add/subtract,                                 
at least the operands won't get messed up.                                                          
Tests on line 2620: If both are true, then Paranoia presumes chopping                                 
1-U1*U1 -> to 1, (false)                                                                            
1+U2*(0.5-U2) = 1+0.5*U2 - U2^2 -> 1+1*U2 (false)                                                 
Tests on line 2650.  Failure means not correctly rounded.                                             
1+(0.5+U2)*U2 = 1+0.5*U2 + U2^2 -> 1+U2 (correct)                                                   
1+(0.5-U2)*U2 = 1 +0.5*U2 -U2^2 -> 1 (correct)                                                    
Tests on line 2670.  Failure means not correctly rounded.                                             
1-(0.5+U2)*U1 = 1 - 0.5*U1 - U2*U1 -> F9 (correct)                                                  
1 - (0.5-U2)*U1 = 1 - 0.5*U1 + U2*U1 -> 1-1*U1, (not correct)                                     
S = (X+Y) + (Y-X) == 0 test (line 2720): passed                                                     
SUBTRACTION(less MACs): Is neither chopped nor correctly rounded                                                                                                                                                                                                                                            

TESTING X*Y == Y*X                                                                                  
MULTIPLICATION: 10 pairs commuted                                                                                                                                                                                                                                                                           

----BEGINNING NON-PARANOIA TESTS----                                                                                                                                                                                                                                                                        

EXTRA MULTIPLICATION TESTS                                                                          
This setup presumes chopping                                                                          
(1+2*U2)(1-2*U2) = 1+4092*U1                                                                        
(1+4*U2)(1-4*U2) = 1+2046*U2                                                                        
(1+2*U2)(1+U2) = 1+4100*U1                                                                          
(1+3*U2)(1+2*U2) = 1+0*U1                                                                                                                                                                             

EXTRA ADDITION/SUBTRACTION TESTS                                                                    
This setup presumes chopping                                                                       
1-U1 < 1                                                                                            
1-F9*U1 = F9                                                                                        
1-0.5*U1 = 1                                                                                        
1-F9*0.5*U1 = 1                                                                                     
SUBTRACTION: Appears to have guard bit, no round bit, and rounds operandsup
	     at least sometimes (?]                                                                                                                                                                                                          

----BEGINNING ERROR MEASUREMENT TESTS----                                                                                                                                                               
Testing Multiply Combinations                                                                       
Exhaustive Sets progress:                                                                           
....................................                                                                
Error bound for mult= [-0.78125,0.625] ULPs                                                                                                                                                             
Testing Division Combinations                                                                       
Exhaustive Sets progress:                                                                           
....................................                                                                
Error bound for div= [-1.19902,1.37442] ULPs                                                                                                                                                            
Testing Subtraction Combinations                                                                    
Exhaustive Sets progress:                                                                           
....................................                                                                
Error bound for sub= [-0.75,0.75] ULPs                                                                                                                                                                  
Testing Addition Combinations                                                                       
Exhaustive Sets progress:                                                                           
....................................                                                                
Error bound for add= [-1,0] ULPs                                                                                                                                                                        
----------------------------------------------                                                          
SUMMARY                                                                                         
----------------------------------------------                                                      
Profile = fp40                                                                                      
Readback seems to work                                                                              
U1 U2 F9 B9 Verified
                                                                                                                                                                                    
----BEGINNING PARANOIA TESTS----                                                                    
No fuzziness detected in comparison                                                                 
MULTIPLICATION: Seems to have guard bit                                                             
MULTIPLICATION: Accuracy tests passed                                                               
DIVISION: FAILURE: Division lacks guard digit                                                       
SUBTRACTION: Seems to have guard bit                                                                
SUBTRACTION: Comparison and Subtraction consistent (2170)                                           
MULTIPLICATION: Is neither chopped nor correctly rounded                                            
DIVISION: Is neither chopped nor correctly rounded                                                  
SUBTRACTION: Is neither chopped nor correctly rounded                                               
SUBTRACTION: (X-Y) + (Y-X) == 0 test passed                                                         
SUBTRACTION(less MACs): Is neither chopped nor correctly rounded                                    
SUBTRACTION(less MACs): (X-Y) + (Y-X) == 0 test passed                                              
MULTIPLICATION: 10 pairs commuted                                                                                                                                                                       

----BEGINNING NON-PARANOIA TESTS----                                                                
SUBTRACTION: Appears to have guard bit, no round bit, and rounds operandsup
	     at least sometimes (?]

----BEGINNING ERROR MEASUREMENT TESTS----                                                           
Error bound for mult= [-0.78125,0.625] ULPs                                                         
Error bound for div= [-1.19902,1.37442] ULPs                                                        
Error bound for sub= [-0.75,0.75] ULPs                                                              
Error bound for add= [-1,0] ULPs
\end{verbatim}

  }
}

\end{document}